\documentclass[preprint,12pt]{elsarticle}

\usepackage{amsmath}
\usepackage{graphicx,psfrag,epsf}
\usepackage{enumerate}
\usepackage[numbers]{natbib}
\usepackage{algorithm}
\usepackage{url}
\usepackage{array}
\usepackage{algorithmicx}
\usepackage[noend]{algpseudocode}
\usepackage{soul}
\algnewcommand\algorithmicinput{\textbf{INPUT: }}
\algnewcommand\Input{\item[\algorithmicinput]}
\algnewcommand\algorithmicoutput{\textbf{OUTPUT: }}
\algnewcommand\Output{\item[\algorithmicoutput]}

\newcommand{\blind}{0}

\journal{Epidemics}

\begin{document}

\begin{frontmatter}



\title{Evaluating COVID-19 Surveillance Testing Strategies at Colorado School of Mines: A Stochastic Modeling Approach}


\author[inst1]{Laura Albrecht}
\author[inst2,inst3,inst4]{Karin Leiderman}
\author[inst5]{Suzanne S. Sindi}
\author[inst1]{Douglas Nychka}

\affiliation[inst1]{organization={Applied Mathematics and Statistics Department, Colorado School of Mines },
            city={Golden},
            state={CO},
            country={USA}}

\affiliation[inst2]{organization={Mathematics Department, University of North Carolina, Chapel Hill},
            city={Chapel Hill},
            state={NC},
            country={USA}}
            
 \affiliation[inst3]{organization={UNC Blood Research Center, University of North Carolina, Chapel Hill},
            city={Chapel Hill},
            state={NC},
            country={USA}}

\affiliation[inst4]{organization={Computational Medicine Program, University of North Carolina, Chapel Hill},
            city={Chapel Hill},
            state={NC},
            country={USA}}
\affiliation[inst5]{organization={Department of Applied Mathematics, University of California, Merced},
            city={Merced},
            state={CA},
            country={USA}}

\begin{abstract}
This study introduces a stochastic model of COVID-19 transmission tailored to the Colorado School of Mines campus and evaluates surveillance testing strategies within a university context. Enhancing the conventional SEIR framework with stochastic transitions, our model accounts for the unique characteristics of disease spread in a residential college, including specific states for testing, quarantine, and isolation. Employing an approximate Bayesian computation (ABC) method for parameter estimation, we navigate the complexities inherent in stochastic models, enabling an accurate fit of the model with the campus case data. We then studied our model under the estimated parameters to evaluate the efficacy of different
testing policies that could be implemented on a university campus. This framework not only advances understanding of COVID-19 dynamics on the Mines
campus but serves as a blueprint for comparable settings, providing insights for
informed strategies against infectious diseases. 
\end{abstract}




\end{frontmatter}

\def\spacingset#1{\renewcommand{\baselinestretch}%
{#1}\small\normalsize} \spacingset{1}


\if1\blind
{
  \title{\bf Title}
  \author{Author 1\thanks{
    The authors gratefully acknowledge the Office of Institutional Research at Colorado School of Mines for their funding of this project.}\hspace{.2cm}\\
    Department of YYY, University of XXX\\
    and \\
    Author 2 \\
    Department of ZZZ, University of WWW}
  \maketitle
} \fi

\bigskip
\begin{abstract}
\end{abstract}

\noindent%
\vfill

\newpage
\spacingset{1.45} 

\section{Introduction}
The COVID-19 pandemic affected society and public health decision making in many ways. One notable example is how this epidemic was managed on college campuses. In most cases, campus administrators had a degree of autonomy in implementing policies regarding regular testing, isolation of those exposed, and quarantine of infected individuals. The unique nature of college environments, characterized by communal living spaces, extensive contact networks, and shared facilities, significantly increases the risk of disease transmission. Following the widespread shift to remote learning in March 2020, there was a national push to resume in-person education due to its benefits for learning outcomes and financial pressures \citep{paxson2020college}. Consequently, extensive research has been conducted to model the spread of COVID-19 within college settings, in order to identify the most effective strategies for a safe reopening such as employing non-pharmaceutical interventions such as  wearing masks, social distancing, and testing to mitigate spread  \citep{borowiakControllingSpreadCOVID192020, brownSimpleModelControl2021, paltielAssessingCOVID19Prevention2021, schultesCOVID19TestingCase2021, foxResponseCOVID19Outbreak2021, bahl2021modeling, gressman2020simulating, paltiel2020assessment}.  Among the strategies explored, surveillance testing emerged as a critical policy tool. For example, (\citet{gressman2020simulating}) employed agent-based models to evaluate the impact of various control measures in reducing virus spread within a large residential university. They found that moving large classes online and increasing test sensitivity were the most effective strategies. Similarly, \citet{paltiel2020assessment} explored cost-effective reopening strategies, finding that testing every two days, even with tests of low sensitivity, could significantly limit the size of outbreaks on a college campus.

A residential campus can be considered a quasi-isolated environment, this provides the opportunity to model the dynamics of COVID-19 in a small and largely closed population. Our goal is to develop a stochastic model that mirrors COVID-19's spread on the Colorado School of Mines (Mines) campus and to fit the model to campus case data.  We develop an extended SEIR model with stochastic transitions to represent the dynamics of the Mines campus and employ an Approximate Bayesian Computation (ABC) method for fitting to campus case data. Using the collected posterior samples and credible intervals, we then forward simulate the model under different testing strategies and scenarios to predict the impact of these changes on the trajectory of the disease. This allows us to investigate the potential effect of interventions, such as changes in testing frequency or number of tests, on the spread of the disease. We use this model to predict the expected number of cases and the timing of future outbreaks with associated uncertainty. 

While many studeies have modeled COVID-19 on college campuses, our approach stands out due to its focus on stochastic modeling. This choice allows for a more accurate variation of disease transmission in a moderate-sized population compared to traditional differential equation models. Incorporating stochasticity is particularly important when the initial number of infections is small and variability in transmission significantly influences the trajectory of the epidemic, as highlighted in \citet{allenPrimerStochasticEpidemic2017}. Stochastic models are also favored for their superior ability to quantify prediction uncertainties \cite{kingAvoidableErrorsModelling2015}. For parameter estimation, we adopt an approximate Bayesian computation (ABC) approach, a method that provides an approximation of the posterior distribution by simulating the model under conditions where direct likelihood computation is challenging or impossible \citep{beaumont2002approximate, mckinley2009inference}. This strategy not only ensures a precise alignment of our model with observed data but also allows for a comprehensive assessment of predictive uncertainties.  Finally, being focused on a single university and part of the Mines COVID task force, we are able to address specific issues that would help to inform decision making by the Mines President and Provost. This kind of analysis may be helpful in the future if Colorado, and Mines, in particular are again beset by a pervasive epidemic.

This paper is structured as follows. In Section 2 we introduce the CSM COVID-19 case data. We also present the details of our extended compartmental model, including our methodology for fitting the model with campus case data. Furthermore, we outline the various testing strategies we simulate, utilizing our calibrated model to assess their impacts. Section 3 presents the outcomes of fitting our stochastic model to the data and the results from simulating different testing strategies. We conclude with a discussion of our findings, highlighting both the strengths and the limitations of our methodology.

\section{Methods}
\subsection{Data}

The Colorado School of Mines is a public research university located just west of Denver in Golden, Colorado. The campus consists of approximately 5,000 undergraduate and 1,500 graduate students. On March 13, 2020 in response to the COVID-19 pandemic, like many universities, the Colorado School of Mines transitioned to fully remote for the rest of the Spring semester. In Fall of 2020, students were given the option of returning to campus with 70\% of classes offered in an in-person or hybrid format. For the safety of the students and faculty returning to campus, Colorado School of Mines partnered with COVIDCheck Colorado (CCC), a social benefit enterprise of the Gary Community Investment Company, to implement a COVID-19 surveillance testing program on the Mines campus \cite{COVIDCheckColorado}. Results were guaranteed to be returned within 48-72 hours.  All students living on campus, athletes, and some face-to-face faculty were included in the surveillance testing pool and were being tested every two weeks.  Any student exhibiting symptoms could be tested either through CCC or though the Student Health Center. Campus administrators were interested in examining the effect of different potential surveillance testing strategies on the spread of the disease on campus.  Specifically, how often should testing of students be conducted and what is the impact  of students who, for various reasons, fail to be tested regularly. 

\begin{figure}[!htbp]
    \centering
\includegraphics[width=\textwidth,height=\textheight,keepaspectratio]{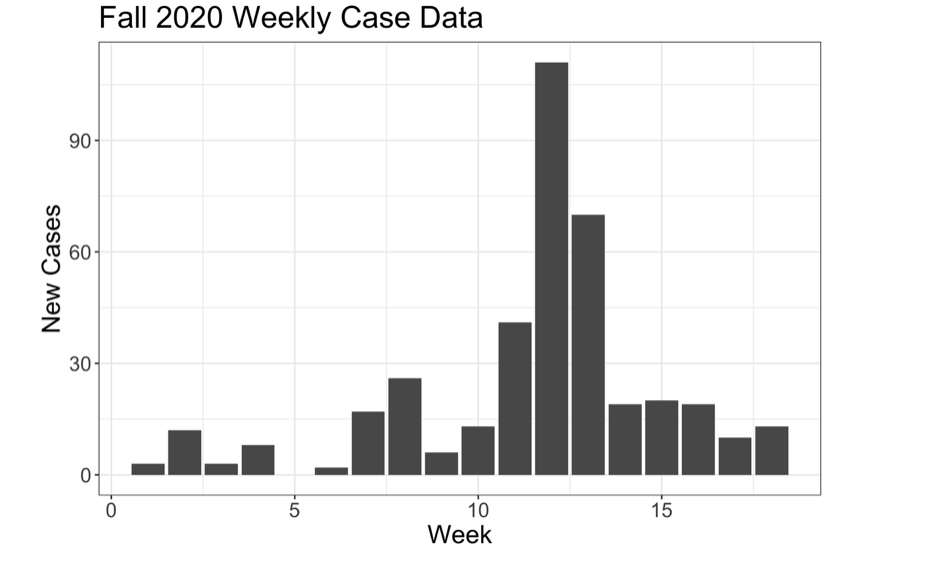}
    \caption{Weekly new COVID-19 cases reported among Colorado School of Mines students during the Fall 2020 Semester. }
    \label{fig:cases}
\end{figure}

\begin{figure}[!htbp]
    \centering
    \includegraphics[width=\textwidth,height=\textheight,keepaspectratio]{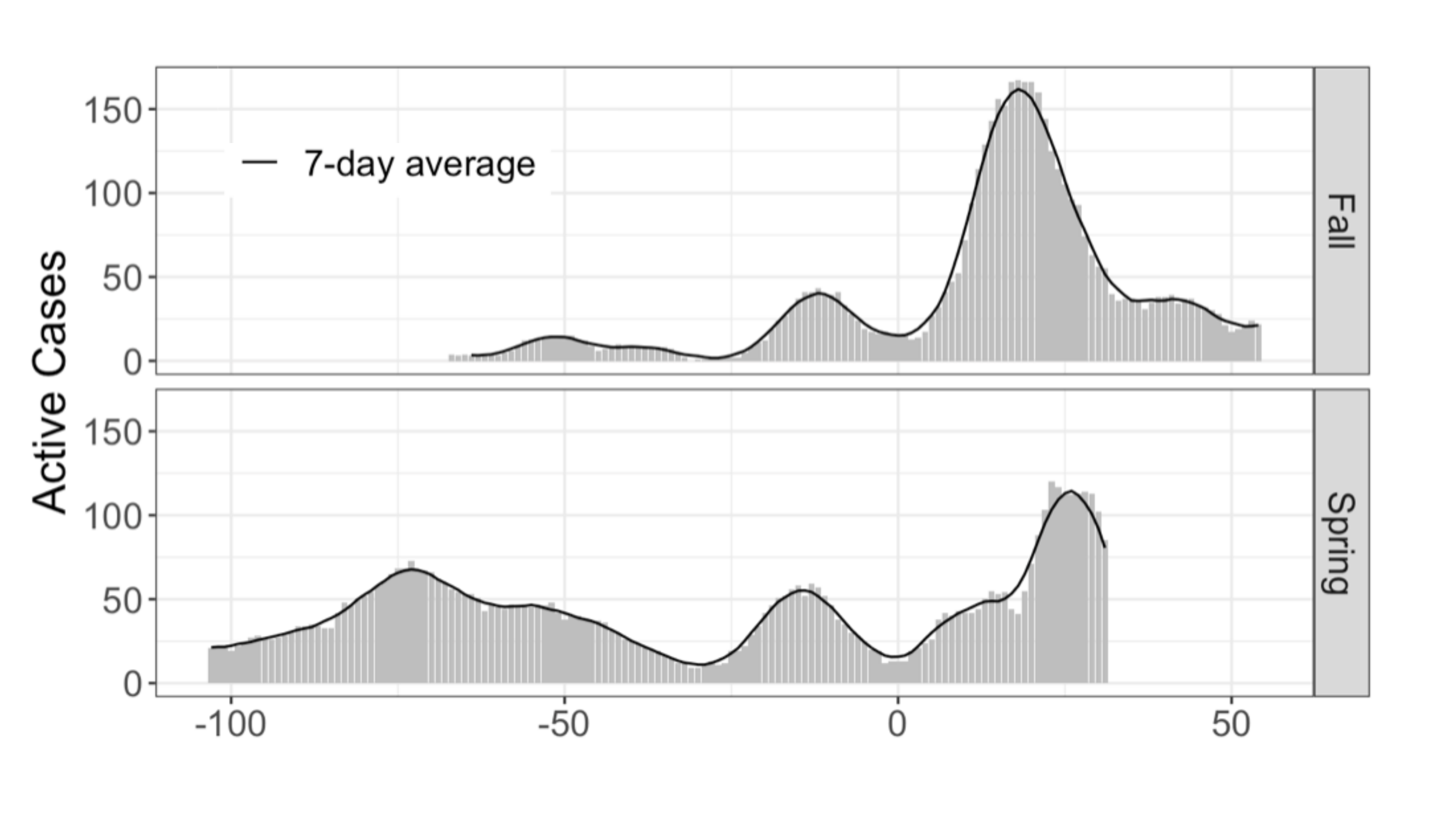}
    \caption{Dynamics of Active Daily COVID-19 Cases Centered Around Fall and Spring Breaks. Active daily cases, calculated from the time of a positive test until the end of a student's 10-day isolation period, are shown in gray while the solid black line represents the centered rolling 7-day average. Fall break occurs at day 70 in the semester, while spring break occurs on day 105. Both semesters exhibit similar patterns: an initial surge of cases at the start of the semester, followed by a second surge just before the break, and the largest surge occurring after the break }
    \label{fig:centered}
\end{figure}

Campus case data for the Fall 2020 semester are shown in Figure \ref{fig:cases}. We observe a small peak in cases in week 8 and a larger surge in cases in weeks 12-13. Similar trends are present in the timing of peaks in cases in the spring semester relative to the school breaks (Figure \ref{fig:centered}). In each semester, we observe multiple peaks of increasing intensity in cases. This cannot be captured with a classical and deterministic SEIR model as infections will always increase quickly at the beginning of an outbreak and subsequent waves will be of a lower intensity if assuming fixed rates for all parameters. We hypothesize the peaks that occur after school breaks are due to students experiencing an increased rate of infection from off-campus travel or other activities.

Campus administrators have a crucial goal of mitigating the peak number of COVID-19 cases on campus through testing. Our study centers around the hypothesis that these peaks in cases can be effectively controlled through a strategic testing approach. We investigate the impact of various surveillance testing strategies on the spread of the disease within the campus community, with a particular focus on the frequency of testing and the number of students being tested. We aim to provide valuable insights that can allow campus administrators to make informed decisions to safeguard the health and well-being of students and faculty at the Colorado School of Mines.

\subsection{Model}
 
We first describe a simple SEIR model and then propose an extended model with additional compartments to more accurately represent the Mines campus. An SEIR model is a common compartmental model used in epidemiology to track the spread of a disease through a population \citep{brauer2008mathematical}. This model represents movement of individuals between stages of the disease; Susceptible, Exposed, Infected, and Recovered. A {\it susceptible} individual is someone who can become infectious. After contact with an infectious individual, a susceptible individual may transition to the {\it exposed} compartment. Exposed individuals have been infected with the virus but are not yet contagious or detectable through testing. {\it Infected} individuals are those currently infectious with the disease and able to spread the disease.  Finally {\it recovered} individuals are those that have previously had the disease and are now recovered. Recovered individuals are assumed to be immune from becoming reinfected. Individuals move among these compartments at different rates depending upon the specific disease. Often, membership is represented mathematically as the proportion of individuals in a given compartment using a system of ordinary differential equations (ODEs) as described in Equations \ref{eq:seir_ode1} - \ref{eq:seir_ode4}:

\begin{align}
     \frac{dS}{dt} &= -\frac{\beta  I}{N}S,
     \label{eq:seir_ode1}
    \\
    \frac{dE}{dt} &= \frac{\beta I}{N}S - \mu E,
    \\
    \frac{dI}{dt} &= \mu E - \gamma  I,
    \\
     \frac{d R}{d t} &= \gamma  I.
     \label{eq:seir_ode4}
\end{align}

For small populations this continuous and deterministic representation is not accurate as the actual amount of contact between individuals may have significant variability \citep{kingAvoidableErrorsModelling2015}. Also the use of continuous time is an approximation since the dynamics of COVID-19  in an actual population has a natural time scale at a daily level with finer granularity involving more detailed  diurnal features of individual behavior.  

Incorporating stochasticity into our model, we evolve the system daily using difference equations. This approach is encapsulated in what is termed a {\it chain binomial model}, a discrete-time stochastic counterpart to traditional continuous-time models, initially introduced by \citet{bailey1975j}. This model introduces a stochastic element by assuming that transitions of individuals between compartments at each time step follow a binomial distribution.  The key  feature is that the  probabilities in the binomial distribution are equal to the rates used in Equations \ref{eq:seir_ode1} - \ref{eq:seir_ode4}.  For instance, the transition from compartment $S$ to compartment $E$ at time $t+1$ is represented by $n_{SE} \sim Bin(S(t), \frac{\beta I(t)}{N})$, where $n_{SE}$ denotes the number of susceptible individuals that have been exposed at time $t$. Similarly, all other transitions follow the same rules. Thus, the difference equations are formulated as Equations \ref{eq:seir_diff1} - \ref{eq:seir_diff4}. In this way our stochastic version will recover the deterministic difference equations from the SEIR as the population size becomes very large and the transitions in continuous time are well approximated by daily differences. 

\begin{align}
     S(t +1) &= S(t) - n_{SE}      \label{eq:seir_diff1}
     \\
     E(t + 1) &= E(t) + n_{SE} - n_{EI}\\
     I(t+1) &= I(t) + n_{EI} - n_{IR} \\
     R(t+1) &= R(t) + n_{IR}
     \label{eq:seir_diff4}
\end{align}

\begin{figure}[!htbp]
    \centering
    \includegraphics[width=\textwidth,height=\textheight,keepaspectratio]{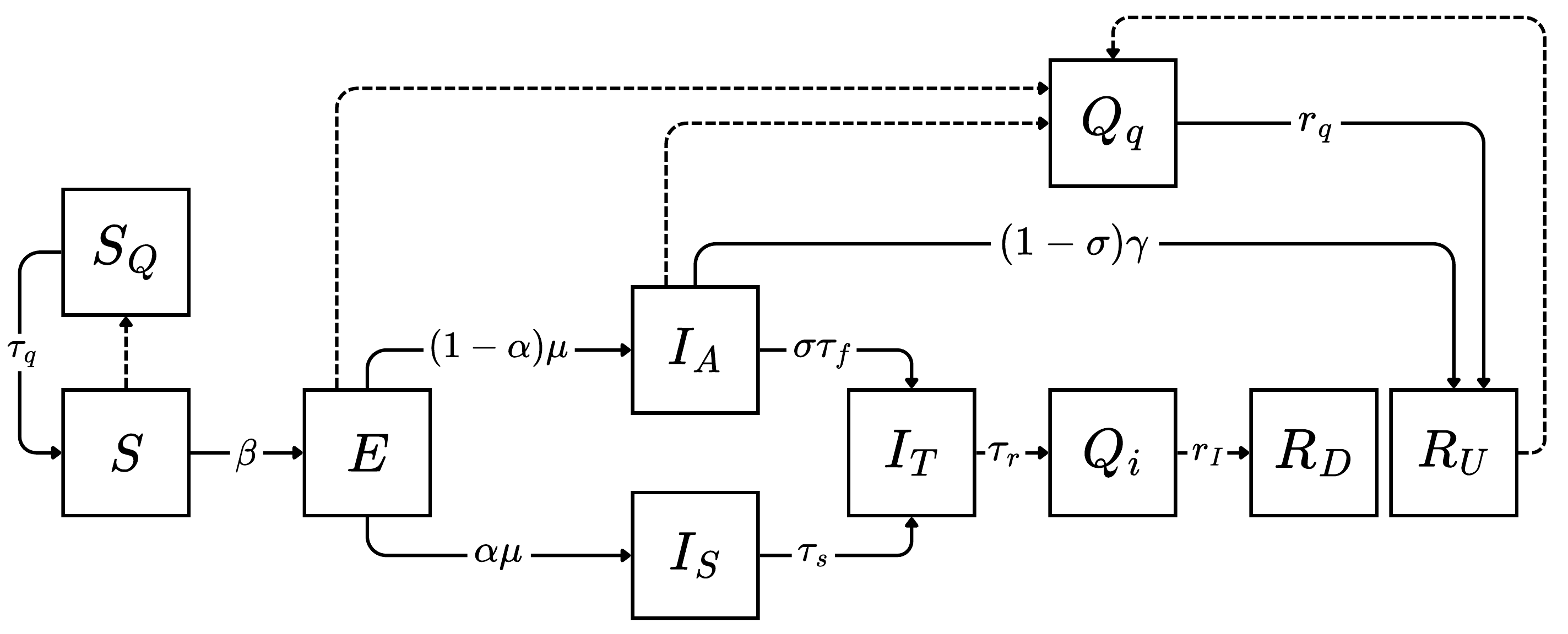}
    \caption{Colorado School of Mines extended SEIR model. Flowchart depicting the dynamic interactions and transitions among individual students through different compartments in an extended SEIR model. The model tracks individuals on campus through stages of COVID-19, including compartments for Susceptible ($S$), Susceptibles in quarantine ($S_q$), Exposed ($E$), Asymptomatic ($I_a$), Symptomatic ($I_s$), Infectious individuals being tested ($I_t$), Isolation($Qi$), Quarantine ($Q_q$), Recovered Detected ($R_d$), and Recovered Undetected ($R_u$). The arrows represent the flow of individuals between compartments, with respective parameters governing the transitions as detailed in \ref{Tab:Parameters1}. }
    \label{fig:flow}
\end{figure}

The basic SEIR model is extended to  include compartments specific to both COVID-19 and the Colorado School of Mines campus. These extensions  are itemized below and the complete model is  depicted in \ref{fig:flow}. 
Some of these compartments facilitate tracking specific populations and numbers of cases to aid with campus policies.   For example, the inclusion of this testing compartment allows us to constrain the total number of tests being administered and evaluate the impact of increasing our capacity or frequency of surveillance testing. Additionally, it enables us to measure the impact of longer or shorter waiting times for results on the spread of the disease.

\subsubsection*{Extended Compartments:}
\begin{itemize}

\item We added separate quarantine ($S_Q$, $Q_q$) and isolation ($Q_i$) compartments due to limited campus space allocation, and tracking the number of students in quarantine/isolation holds high importance for campus administrators.

\item We included two infectious compartments to differentiate between symptomatic and asymptomatic infections, $I_S$ and 
$I_A$ respectively. 

\item We introduced a compartment to explicitly model asymptomatic individuals who become infectious but remain untested throughout their infectious period. These individuals transition directly to the Recovered Undetected compartment ($R_U$) at a rate denoted by $\gamma$.

\item We integtated a testing compartment ($I_T$) to monitor infectious individuals who have undergone testing. Those in the surveillance testing group receive regular tests at a rate of $\tau_f$, with the proportion of students in the surveillance testing program represented by $\sigma$. Symptomatic individuals are tested after developing symptoms at a rate of $\tau_s$. All infectious individuals who undergo testing transition to the $I_T$ compartment until their results are received. For simplicity, we assume test results are 100\% accurate.

\end{itemize}

The role of isolation and quarantine for the campus population is an important aspect for the dynamics of this system and of interest to administrators making decisions. For these reasons, more detail is given on how this is handled in the model. Individuals that receive a positive test will transition into Isolation, $Q_i$. Individuals in Isolation will no longer be able to spread the disease to others and will eventually transition into Recovered Detected, $R_D$. For every individual that enters isolation, some number of their close contacts, $n_{CC}$, will enter into quarantine ($Q_q)$. These $n_{CC}$ close contacts will be chosen from the eligible quarantine compartments, $S, E, I_A, I_S, R_U$. Individuals who have previously tested positive and are now in $R_D$ are temporarily exempt from quarantine. We assume a ``perfect" quarantine in which individuals who enter quarantine as susceptible will not become infectious during their quarantine period. Additionally, exposed or infectious individuals in quarantine will not spread the disease to anyone else outside of quarantine. We denote the susceptible individuals in quarantine as $S_q$ and all other individuals in quarantine as $Q_q$. Susceptible individuals in quarantine will transition back to $S$ at the end of their quarantine period. Individuals in $Q_q$ will transition to either $R_D$ or $R_U$ at the end of their quarantine period.

Our complete model is formulated as in Equations \ref{eq:seir_full1} - \ref{eq:seir_full11}. Transition probabilities and parameters are listed in Table \ref{Tab:Transitions} and Table \ref{Tab:Parameters1} respectively. We assume all parameters are fixed except for the transmission rate, $\beta$, and the proportion of infections that are symptomatic, $\alpha$ and an $I_{\text{out}}$ an increase in exposed students returning from a break. The ranges on $\alpha$ and $\beta$ coincide with reasonable estimates for $R_0$ ranging from 0.8 to 14.8, as we will discuss in the next section. The fixed parameters are set at either values found in the literature on the phases of the pandemic \citep{byrne2020inferred, lauer2020incubation} or known values specific to our campus. Individuals transitioning from each compartment are again treated as binomial draws with the exception of individuals transitioning into testing or quarantine compartments which depend on the size of the surveillance testing pool and the number of individuals that test positive, respectively. Additionally, we assume that after Fall/Spring break, some students will become exposed off-campus and will return as Exposed after the break. We denote the number that become infected during the break as $I_{\text{out}}$. Thus, we have three unknown parameters in the model that we will fit with the campus case data: $\alpha, \beta, \text{and } I_{\text{out}}$. All analyses were performed using \textsf{R} Statistical Software (4.3.1, R Core Team 2023) \cite{R}.

\begin{table}[!htbp]
\begin{tabular}{|c|c|c|c|c|c|c|c|c|c|c|c|c|}
\hline
\centering
\textbf{From} & $S$ & $E$ & $E$ &$I_A$ & $I_A$ &$I_S$ &$I_S$ & $I_T$ & $Q_i$ & $Q_q$ & $S$ & $E, I_A, R_U$ \\ \hline
\textbf{To} & $E$ & $I_A$ & $I_S$ & $I_T$ & $R_U$& $I_T$ & $Q_i$ & $R_U$&   $R_D$ & $R_U$ & $S_q$ & $Q_q$\\ \hline
\textbf{Probability} & $\frac{\beta I}{N}$  & $(1- \alpha) \mu$ & $\alpha \mu$ & $\sigma\tau_f$ & $(1- \sigma)\gamma$  & $\tau_S$ & $\gamma$& $\tau_r$  & $r_I$ & $r_q$ & $\tau_q$ & $\tau_q$\\  \hline
\end{tabular}
\caption{Transition Probabilities. Individuals in the first column are selected with the given probabilities to transition to the compartment in the second column compartment at every time period $t$.}
\label{Tab:Transitions}
\end{table}

\begin{table}[!htbp]
\begin{tabular}{|c|c|c|c|c|c|}
\hline
$\mathbf{Parameter }$                                        & $\mathbf{Description  }$           & $\mathbf{Estimate }$  & $\mathbf{Fixed}$ & $\mathbf{Est}$ & $\mathbf{Policy}$    \\ \hline \hline
$\beta$  &  Transmission Rate                                 &            0.2 - 1 (0.4)  & & X &        \\ \hline
$\alpha$ & Proportion symptomatic                                   & 0.2 - 0.8 (0.3)  & & X &\\ \hline
$1/\mu$ &  Duration exposed but not infectious     & 3 days & X &  &             \\ \hline 
 
 $1/\gamma$      & Duration detectable and infectious               &14 days     & X & &          \\ \hline 
 $\sigma$          & Proportion students in surveillance testing                   & 0-1 (0.4)  &  & &X \\ \hline
 $1/\tau_f$   & Surveillance testing frequency                    & 3-14 days (14)       & & & X       \\ \hline
  $1/\tau_s$   & Duration symptomatic before testing       & 2 days       & X & &        \\ \hline
$1/\tau_r$   & Waiting time for test results                & 2 days        & X & &       \\ \hline
$1/r_I$      & Duration of isolation                             & 10 days     & X & &         \\ \hline

$1/r_q$  & Duration of quarantine                  & 14 days & & & X \\ \hline
 $N_{cc}$ & Number of close contacts 
& 0-20 (10) & X & &  \\ \hline
$I_{out}$ & Infectious students returning after break & 0-200 (100) & & X & \\
\hline
\end{tabular}
\caption{Parameters in the model and their initial estimates. The $\alpha$ and $\beta$ ranges chosen coincide with estimates for $R_0 = (0.8, 14.8)$. We estimate these parameters by fitting with the Colorado School of Mines COVID-19 case data. The fixed parameters are set at either values found in the literature or known values specific to our campus. The policy parameters are those we can control through policy decisions on campus. }
\label{Tab:Parameters1}
\end{table}


\begin{align}
   S(t + 1)     &= S(t)     - n_{SE}  - n_{SS_Q}                     + n_{S_Q S}  \label{eq:seir_full1} \\
   S_Q(t + 1) &= S_Q(t) - n_{S_Q S}                                   + n_{SS_Q}\\
   E(t + 1)     &= E(t)     - n_{EI_A} - n_{E I_S} - n_{EQ_Q} + n_{SE}\\
   I_A(t + 1)   &= I_A(t)  - n_{I_A I_{T}} - n_{I_AQ_Q} - n_{I_A R_U} + n_{EI_A}\\
   I_S(t + 1)   &= I_S(t)  - n_{I_S I_{T}}  - n_{I_S R_U}         + n_{E I_S}\\
   I_{T}(t + 1)  &= I_{T}(t) - n_{I_{T} Q_i}                               + n_{I_A I_{T}} + n_{I_S I_T} + n_{Q_Q I_T}\\
   Q_i(t + 1)   &= Q_i(t)  - n_{Q_i R_D}                                 + n_{I_{T} Q_i}\\
   Q_Q(t + 1) &= Q(t)     - n_{Q_Q I_T} - n_{Q_Q R_U}.        + n_{EQ_Q} + n_{I_AQ_Q} + n_{R_UQ_Q} \\
   R_D(t + 1) &= R_D(t)                                                         + n_{Q_i R_D}\\
   R_U(t + 1) &= R_U(t) - n_{R_UQ_Q}                                + n_{I_A R_U} + n_{I_S R_U} + n_{Q_Q R_U} \label{eq:seir_full11}
\end{align}
where all compartments are non-negative integers for for all $t$.

\subsection{Basic Reproduction Number, $R_0$}
\label{sec_r0}
The basic reproduction number, $R_0$, is defined as the average number of secondary infections when an infected person enters an entirely susceptible population and is a key metric determining the severity of transmission of a disease. \citet{brauer2008mathematical} finds the value of $R_0$ is the same in discrete-time models as in continuous-time models. To explore the relationship between $\alpha$ (the proportion symptomatic) and $\beta$ (the transmission rate) in the model, we find the estimate of $R_0$ using the continuous-time ODE model. The calculation is performed using the next generation matrix method, detailed in the supplemental materials, allowing us to derive an explicit equation for $R_0$, thus providing insights into how changes in model parameters influence disease spread dynamics.

We find the following equation for $R_0$ at the parameter values given in Table \ref{Tab:Parameters1}:


$$ R_0 = (14.8 - 10.8\alpha)\beta$$.

\noindent The proportion of symptomatic individuals, denoted as $\alpha$, significantly influences the
expected basic reproduction number, $R_0$. This is due to our assumption symptomatic individuals are more likely to get tested and, therefore, enter isolation, reducing their role in the transmission cycle. If all individuals are asymptomatic, $R_0$ would increase by a factor of 3.7 compared to a scenario where all individuals are symptomatic.

\begin{figure}[!htbp]
\includegraphics[width=\textwidth,height=\textheight,keepaspectratio]{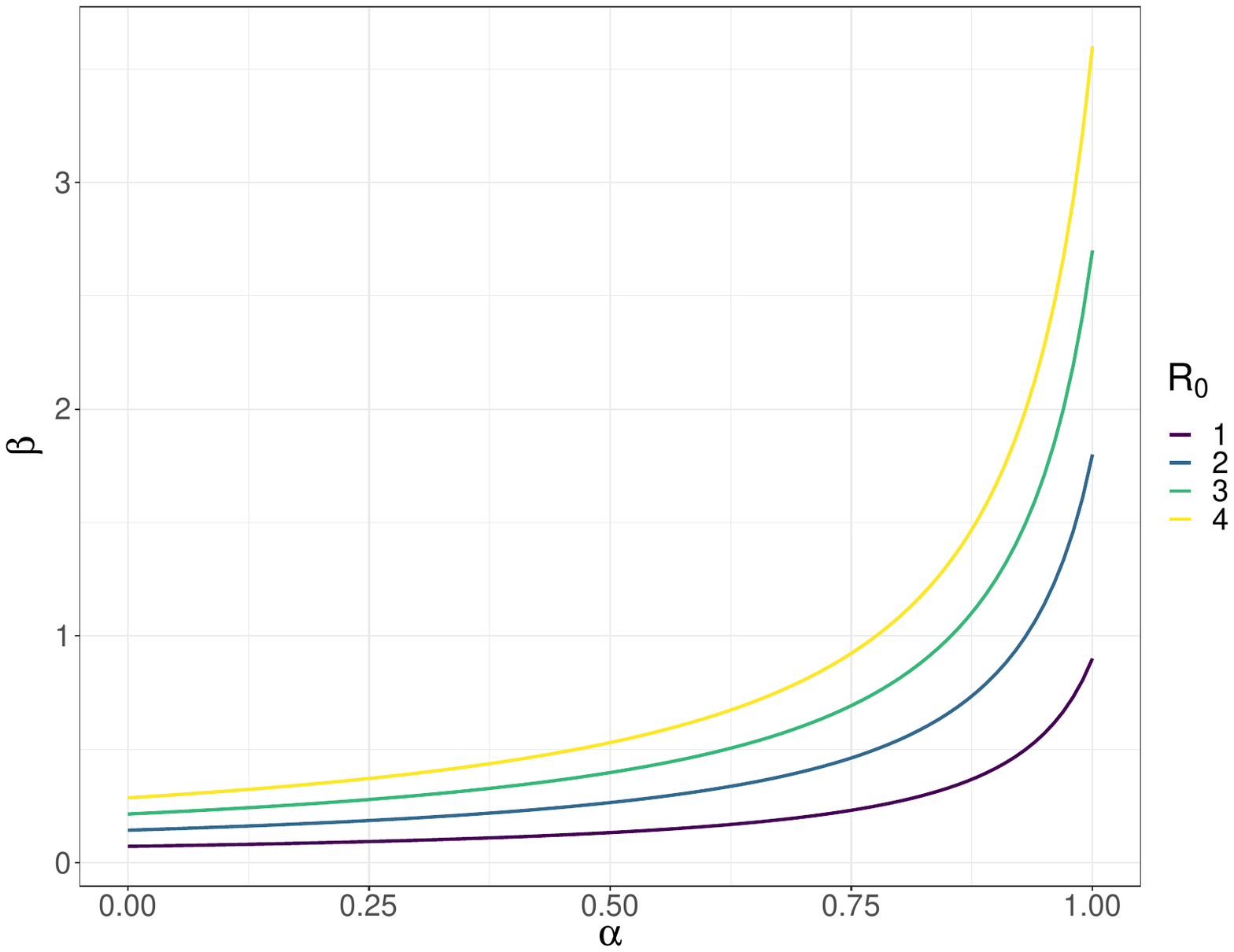}
    \caption{Varying of the Basic Reproductive Number ($R_0$) with Changes in Transmission Rate ($\beta$) and Proportion of Symptomatic Individuals ($\alpha$). This plot depicts the relationship between the transmission dynamics of the disease, characterized by the rate of transmission ($\beta$) and the proportion of symptomatic cases ($\alpha$), and how these two factors jointly influence $R_0$, a key epidemiological metric. See Section \ref{sec_r0} for more details}
    \label{fig:next_gen}
\end{figure}

Figure \ref{fig:next_gen} shows an example of the relationship between $\beta$ and $\alpha$ with $R_0 = 1.0, 2.0, 3.0, 4.0$. This illustrates the dynamic relationship between the transmission rate ($\alpha$) and the recovery rate ($\beta$). Given that $\alpha$ represents a proportion, it is observed that $\alpha$ can fluctuate across its entire feasible range in response to minor changes in $\beta$.

\subsection{Approximate Bayesian Computation (ABC)}
\label{sec:abc}

Due to the stochastic nature of both our model output and data, parameter estimation poses a significant challenge. The multitude of possible trajectories generated by a single set of parameter values makes fitting the model to a single curve difficult. To overcome this, we adopt an approach for estimation that focuses on only the most important aspects of the curve for policy makers. Rather than seeking a perfect fit, we focus on fitting only the peaks in case data. This allows us to extract valuable insights from the data on the timing and height of peaks and make accurate predictions of future case surges. Moreover these statistics reflect the level of temporal resolution that is relevant for how campus administrators were able to monitor case count data. 

\subsubsection{ABC statistics}
We define a ``peak" to be an increase in cases for at least 1 week followed by a decrease for 1 week with the minimum number of weekly cases of at least 20.  We find the number of peaks and their timings  in the Mines COVID case count data for the Fall 2020 semester and for clarity we refer to these as the ABC statistics. Specifically our ABC Statistics are defined as the \textit{number of peaks}, the \textit{timing of the peaks}, and the \textit{height of the peaks}.

The ABC approach involves simulating the model at different parameter values and comparing the distribution of simulated ABC statistics with those in the observed data.  In general this comparison also involves setting a tolerance for continuously varying and multivariate statistics for acceptance.  Parameter values that produce similar ABC statistics to the true values have higher posterior probability.  The parameters we allow to vary in the model are $\beta$, the rate of infection, $\alpha$, the proportion of infected individuals that are symptomatic, and $I_{\text{out}}$, the number of infected individuals that return to campus after break. We assign prior distributions to each of these parameters, such that $\beta \sim U(0.2, 1)$, $\alpha \sim U(0,1)$, $I_{\text{out}} \sim U(1, 200)$. These prior ranges lead to potential $R_0$ values ranging from $0.6 - 14.8$ and so provide a wide range of possible  basic reproduction rates. 

 We create a grid of 21 points across the prior range for $\alpha$, $\beta$ and $I_{\text{out}}$ resulting in $21^3$ combinations.  For each combination of parameters, 1000 trajectories are generated using the full stochastic model. The timing and height of each peak in the simulations are calculated and compared to the true data. A parameter combination is accepted if it satisfies the following conditions: the number of peaks matches the true data, the timing of each peak falls within a 1-week range of the true data, and the height of each peak is within 10 cases of the true data. Specifically, we compute posterior estimates for the parameters as detailed in  Algorithm \ref{alg:abc} and use the posterior samples to determine credible intervals for each of the unknown parameters. We note that this is one of the simplest forms of posterior sampling: we generate the ABC statistics and model parameters from the joint distribution of likelihood and uniform priors and then report the conditional sample of parameters where the ABC statistic values  match the observed statistics. For any combination of model parameters simulation of  multiple realizations of the Mines COVID case count time series is fast and can be done in parallel. In our work we typically generate 1000 realizations of the times series for a fixed set of model parameters and so our Monte Carlo uncertainty for the posterior probabilities are based on binomial sampling of 1000 trials. 

 \begin{algorithm}
\caption{Approximate Bayesian Computation (ABC) for Posterior Estimation}
\begin{algorithmic}[1]
\State Define a uniform grid of parameter values for $\alpha$, $\beta$, and $I_{out}$ within the ranges specified by their respective prior distributions. 
\State Create a set $\Theta$ of  all possible combinations of these parameters within the defined grid.
\For{each parameter combination $\theta_i$ in $\Theta$}
    \State Simulate 1000 trajectories of the model using parameters $\theta_i$
    \For{$j = 1$ to $1000$} 
        \State Calculate ABC statistics for the simulated trajectory
        \State Check if the number of peaks is equal to the data
        \State Check if the timing of the peaks are within +/- 1 week
        \State Check if the heights of the peaks are within +/- 10 cases
          \If{All ABC statistics are within the given thresholds}
            \State Add the parameter values $\theta_i$ to the posterior sample
        \EndIf
    \EndFor
\EndFor
\end{algorithmic}
\label{alg:abc}
\end{algorithm}

\subsubsection{Justification of ABC}
By summarizing our data in terms of these ABC statistics, we have condensed a complex and non-Gaussian time series to a handful of pivotal features.  Note that these statistics can be interpreted as  discretizing time to weeks, the number of cases to 10's of cases. The number of peaks is, of course,  already a discrete set. 
The ABC computation for the time series based on this reduction  is  now  an {\it exact}  Bayesian analysis. Moreover the Bayesian posterior is readily computed using Monte Carlo sampling without resorting to Markov chain sampling. In general the ABC approach can be criticized for being subjective in the choice of statistics and acceptance criteria. However, in this application the choice of ``sufficient statistics" is deliberate, reflecting our understanding of the limits of the data and the specific features of COVID transmission and testing needed to  make campus health decisions. In addition, the granularity of the statistics can always be made finer  to address more features in the case counts and will just involve more sampling \cite{marjoram2003markov}.

\subsection{Evaluating COVID Testing Strategies}
Once our model has been fit to the data, , we use these estimated parameters to simulate the effect of different COVID testing policies that are available to the administration. Specifically, campus health administrators were interested in determining if  testing could lower the number of students in quarantine/isolation and the total number of cases by the end of the semester. At the start of the Fall 2020 semester, approximately 40\% of students attending classes in-person were tested every 14 days as part of the surveillance testing program. We simulated the effect of increasing the proportion of students tested to 60\% and 80\% and increasing the frequency of testing to one time per week and two times per week. In each of these simulations, we compare the total number of students in quarantine and the total number of cases at the end of the semester. Each simulation is shown in Table \ref{tab:testing_strategies}.

\begin{table}
\centering
\begin{tabular}{|c|c|}
\hline
\textbf{Proportion Tested} & \textbf{Testing Frequency} \\
\hline
40\% & Every 14 days \\
\hline
40\% & Weekly \\
\hline
40\% & Twice a week \\
\hline
60\% & Every 14 days \\
\hline
60\% & Weekly \\
\hline
60\% & Twice a week \\
\hline
80\% & Every 14 days \\
\hline
80\% & Weekly \\
\hline
80\% & Twice a week \\
\hline
\end{tabular}

\caption{Simulated Testing Strategies. The first row simulates the baseline used for the Fall 2020 semester. We then simulate the effects of both increasing the proportion of students in the surveillance testing group and increasing the frequency of testing.}
\label{tab:testing_strategies}
\end{table}

\section{Results}

\subsection{Simulation study results} 
We simulated data at different parameter values within the prior ranges for $\alpha, \beta$ and $I_0$ in order to evaluate the effectiveness of our ABC parameter fitting method. 1000 curves were simulated from all permutations of $\alpha = \{0.25, 0.75\}, \beta = \{0.32, 0.8\}, \text{and } I_0 = \{100, 150\}$. Of the 1000 simulated curves, we sampled 100 curves from each set of parameter combinations shown in Table \ref{tab:posterior_sim} and assumed that curve was ``true". Posterior samples were collected using the ABC method described in Algorithm \ref{alg:abc}. Results are shown in Table \ref{tab:posterior_sim} with 95\% credible intervals for each parameter. The 95\% credible intervals encompassed the true simulated values for each parameter set. However, the credible intervals for $\alpha$ were notably broad, offering limited insight beyond the initial priors. To further investigate the challenges of identifying $\alpha$, we repeated this analysis while fixing $\beta$ to its true value, which marginally narrowed the range for $\alpha$ but not significantly. For instance, with fixed $\beta = 0.32$ and settings $\alpha = 0.25, I_{out} = 100$, the credible interval for $\alpha$ was $(0, 0.61)$. 

\begin{table}[!htbp]
\centering
\begin{tabular}{|c|c|c|c|}
\hline
\textbf{Data simulated from:}                                          & \multicolumn{3}{l|}{\textbf{95\% Credible Intervals}}                                                                            \\ \hline
\multicolumn{1}{|l|}{\textbf{}}                                        & \multicolumn{1}{c|}{\textbf{$\alpha $}} & \multicolumn{1}{c|}{\textbf{$\beta $}} & \multicolumn{1}{c|}{\textbf{$I_\text{out}$}} \\ \hline
\multicolumn{1}{|l|}{$\alpha = 0.25, \beta = 0.32, I_\text{out} = 100$} & \multicolumn{1}{l|}{(0, 0.96)}          & \multicolumn{1}{l|}{(0.25, 0.44)}      & \multicolumn{1}{l|}{(43, 157)}               \\ \hline
\multicolumn{1}{|l|}{$\alpha = 0.25, \beta = 0.32, I_\text{out} = 150$} & \multicolumn{1}{l|}{(0, 0.99)}          & \multicolumn{1}{l|}{(0.26, 0.47)}      & \multicolumn{1}{l|}{(58, 178)}               \\ \hline
\multicolumn{1}{|l|}{$\alpha = 0.25, \beta = 0.80, I_\text{out} = 100$} & \multicolumn{1}{l|}{(0, 0.82)}          & \multicolumn{1}{l|}{(0.64, 0.90)}      & \multicolumn{1}{l|}{(57, 192)}               \\ \hline
\multicolumn{1}{|l|}{$\alpha = 0.25, \beta = 0.80, I_\text{out} = 150$} & \multicolumn{1}{l|}{(0, 0.89)}          & \multicolumn{1}{l|}{(0.64, 0.90)}      & \multicolumn{1}{l|}{(59, 193)}               \\ \hline
\multicolumn{1}{|l|}{$\alpha = 0.75, \beta = 0.32, I_\text{out} = 100$} & \multicolumn{1}{l|}{(0, 0.99)}          & \multicolumn{1}{l|}{(0.22, 0.41)}      & \multicolumn{1}{l|}{(53, 166)}               \\ \hline
\multicolumn{1}{|l|}{$\alpha = 0.75, \beta = 0.32, I_\text{out} = 150$} & \multicolumn{1}{l|}{(0, 0.99)}          & \multicolumn{1}{l|}{(0.22, 0.41)}      & \multicolumn{1}{l|}{(53, 166)}               \\ \hline
\multicolumn{1}{|l|}{$\alpha = 0.75, \beta = 0.80, I_\text{out} = 100$} & \multicolumn{1}{l|}{(0.14, 1)}          & \multicolumn{1}{l|}{(0.66, 0.89)}      & \multicolumn{1}{l|}{(70, 196)}               \\ \hline
\multicolumn{1}{|l|}{$\alpha = 0.75, \beta = 0.80, I_\text{out} = 150$} & \multicolumn{1}{l|}{(0.22, 1)}          & \multicolumn{1}{l|}{(0.67, 0.89)}      & \multicolumn{1}{l|}{(73, 196)}               \\ \hline
\end{tabular}
\caption{Posterior Credible Intervals for simulated data. We simulated 100 data curves at each of the parameter values given in the first column. Posterior samples were then collected using the method described in Algorithm \ref{alg:abc}. The 95\% credible intervals for each simulated data set are shown here. In all cases the true simulated value falls within the 95\% credible interval, although the credible intervals for $\alpha$ covers nearly its entire feasible range. }
\label{tab:posterior_sim}
\end{table}


\subsection{Model fitting for Fall 2020} 
\begin{figure}[!htbp]
    \centering
    \includegraphics[width=\textwidth,height=\textheight,keepaspectratio]{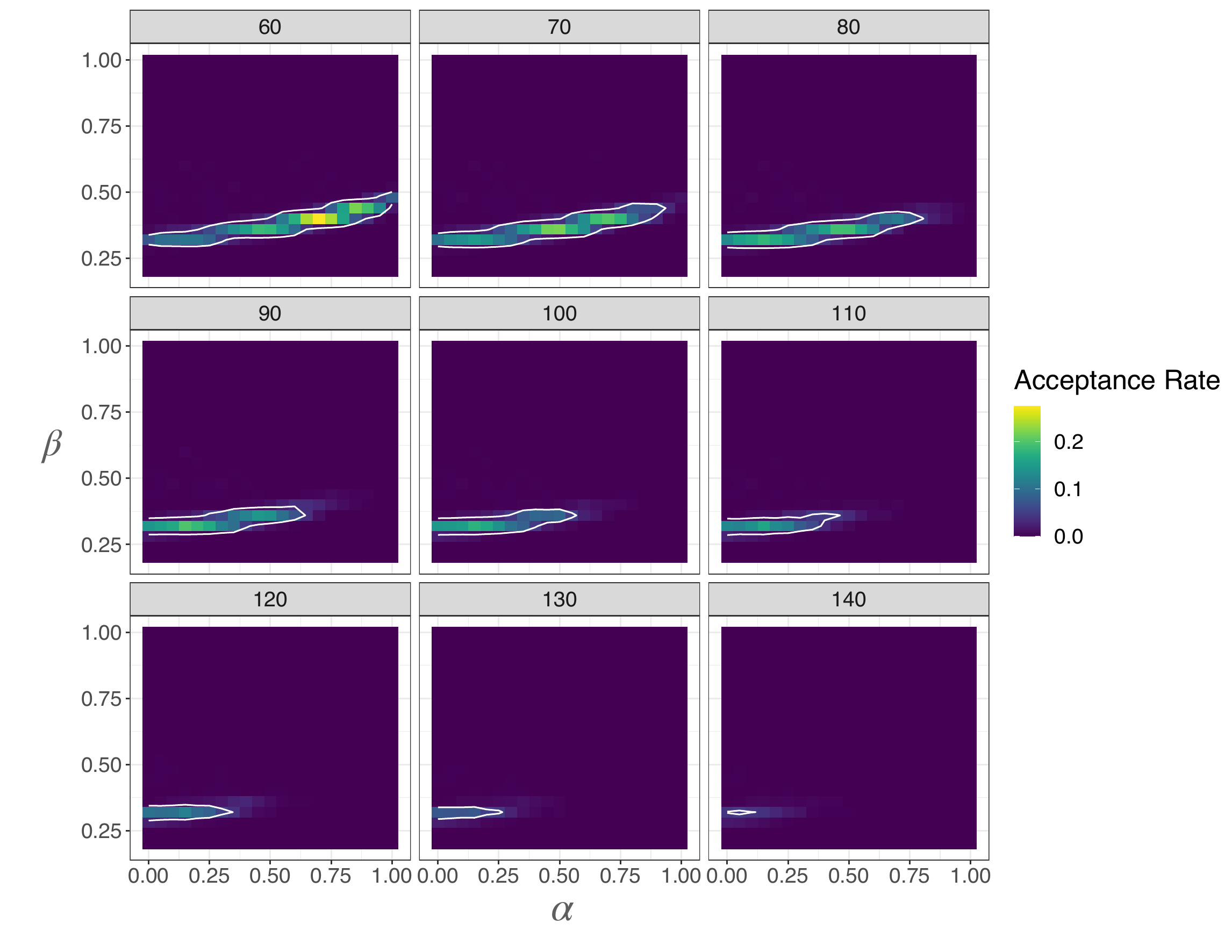}
    \caption{Joint posterior acceptance rates for Fall 2020 data. Each facet label indicates the number of outside infections returning after break with values for $\alpha$ and $\beta$ shown on the x and y axis. The color indicates the proportion of simulations at each given parameter set that were accepted into the posterior sample. The contour line shows the 95\% credible interval at each value of $I_{\text{out}}$. Small changes in $\beta$ lead to large changes in the accepted values for $\alpha$. As the number of outside infections increases, the proportion symptomatic, $\alpha$, decreases to compensate. }
    \label{fig:data_joint}
\end{figure}

\begin{figure}[!htbp]
    \centering
    \includegraphics[width=\textwidth,height=\textheight,keepaspectratio]{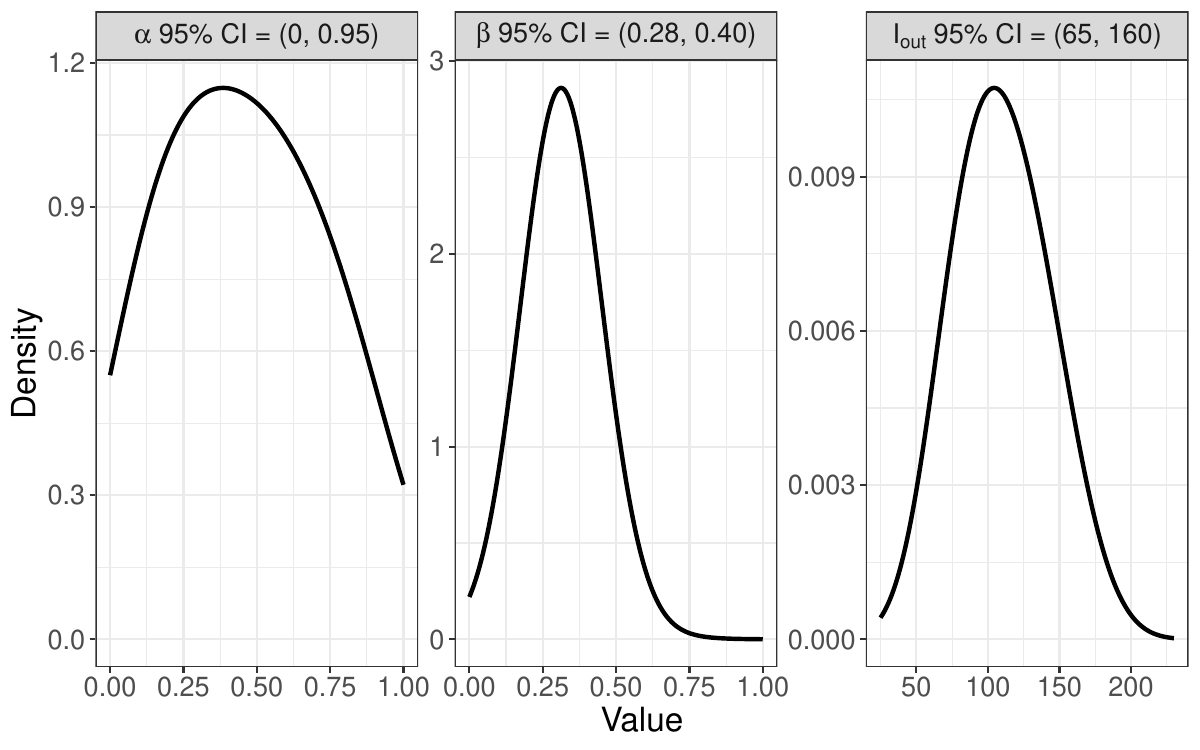}
    \caption{Marginal posterior credible intervals for unknown parameters fitting with Fall 2020 data. The range on $\alpha$, the proportion of individuals symptomatic, is very wide and still mostly undetermined. }
    \label{fig:data_marginal}
\end{figure}

Using the algorithm to determine the model parameters (see Section \ref{sec:abc}) 95\% credible intervals for the campus data from Fall 2020 are found to be $\alpha = (0, 0.95), \beta = (0.28, 0.4), I_{\text{out}} = (65, 140) $. These intervals correspond with estimated $R_0$ values of 1.7 - 3.8 which agree with estimates found in other studies \cite{alimohamadi2020estimate, katul2020global, d2020assessment}. Joint and marginal posterior plots are shown in Figures \ref{fig:data_joint} and \ref{fig:data_marginal} respectively. The range on $\alpha$ is again uninformative and we observe that small changes in $\beta$ lead to large changes in $\alpha$.

\begin{figure}[!htbp]
    \centering
    \includegraphics[width=\textwidth,height=\textheight,keepaspectratio]{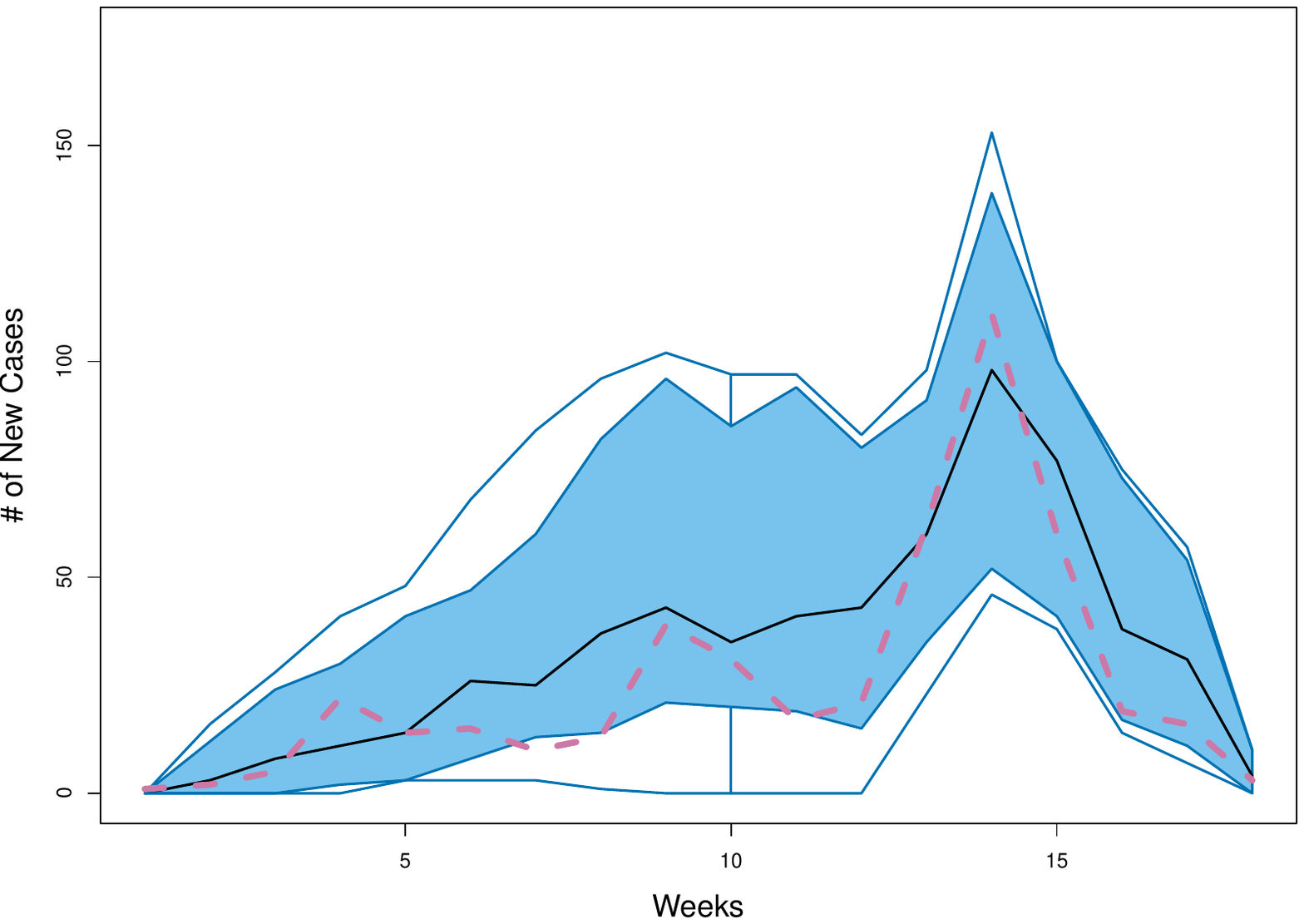}
    \caption{Functional Boxplot of curves simulated from posterior draws. The black curve represents the median, the inner box depicts the 50\% IQR, and the pink curve shows the Fall 2020 case data. The posterior simulations are able to capture the trends and intensity of the peaks in the data.}
    \label{fig:fb_data}
\end{figure}

Based on this posterior distribution for the parameters, an ensemble of 1000 trajectories were simulated. For clarity we refer to these curves as the {\it posterior case count trajectories} and they have the interpretation that each is an equally plausible history conditional on the data and our extended SEIR stochastic model.  Thus, the variation in these trajectories is useful to quantify the amount of uncertainty in the case counts.  Also, if these trajectories exhibit stochastic features very different from the observations it calls into question our model. These curves were summarized by a functional boxplot created in \textsf{R} using the \textsf{fda} package (Figure \ref{fig:fb_data}) \cite{sun2011functional, fda}. The Fall 2020 case data was added for comparison (pink line in 
\ref{fig:fb_data}). 
The posterior trajectories effectively capture the trends in the true data, with the data falling within the 50\% IQR of the simulated curves. The timing and height of the highest peak at week 14 closely align with the median curve.

\subsection{Simulation of different surveillance testing strategies}

\begin{figure}[!htbp]
    \centering
    \includegraphics[width=\textwidth,height=\textheight,keepaspectratio]{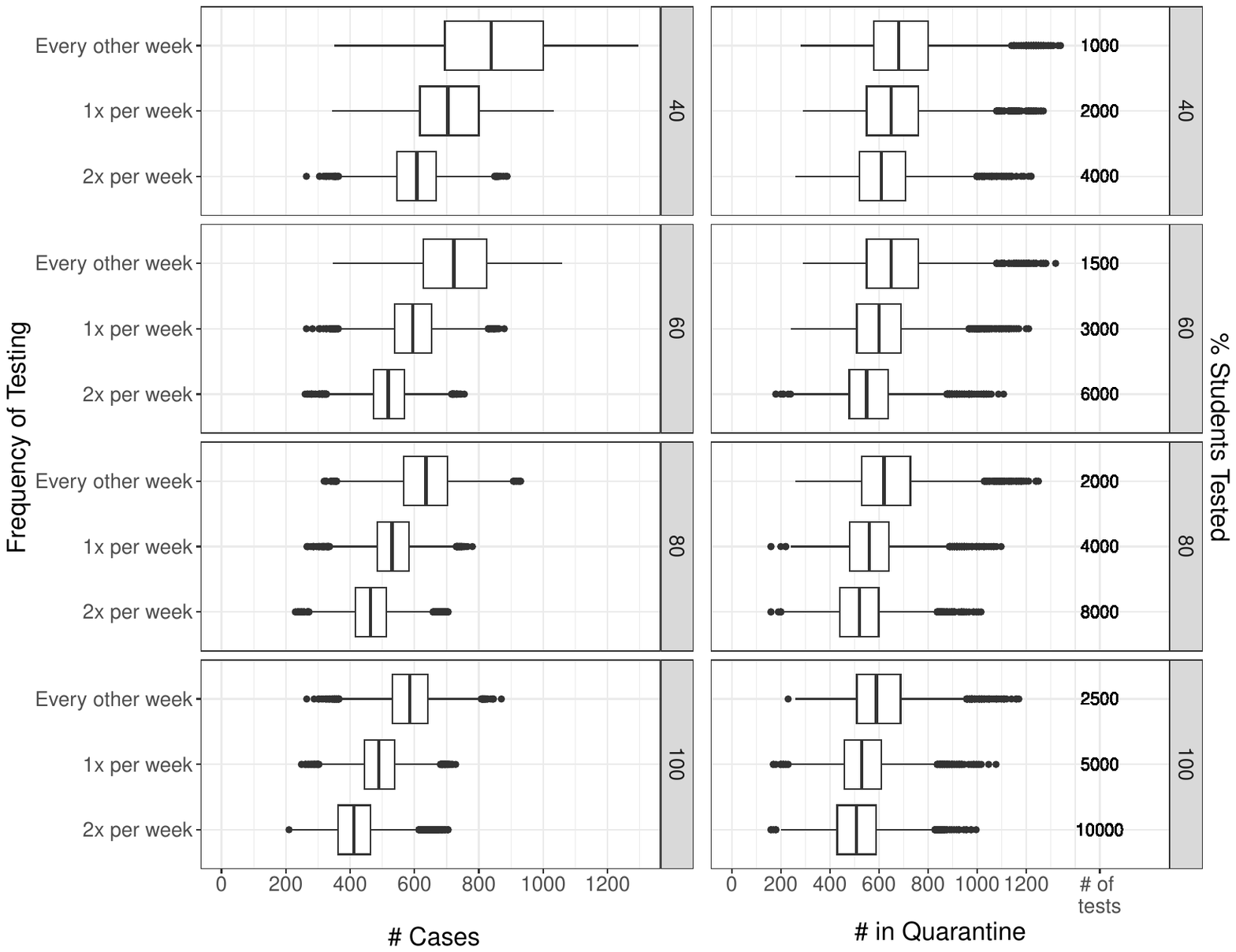}
    \caption{Impact of varying the frequency of surveillance testing and the percentage of students being tested.  Additionally, the total number of tests administered under each scenario is included. While increasing testing frequency and coverage significantly lowers reported cases, it only slightly reduces the number of students in quarantine by semester's end.} 
    \label{fig:freq_prop}
\end{figure}

\begin{figure}[!htbp]
    \centering
    \includegraphics[width=\textwidth,height=\textheight,keepaspectratio]{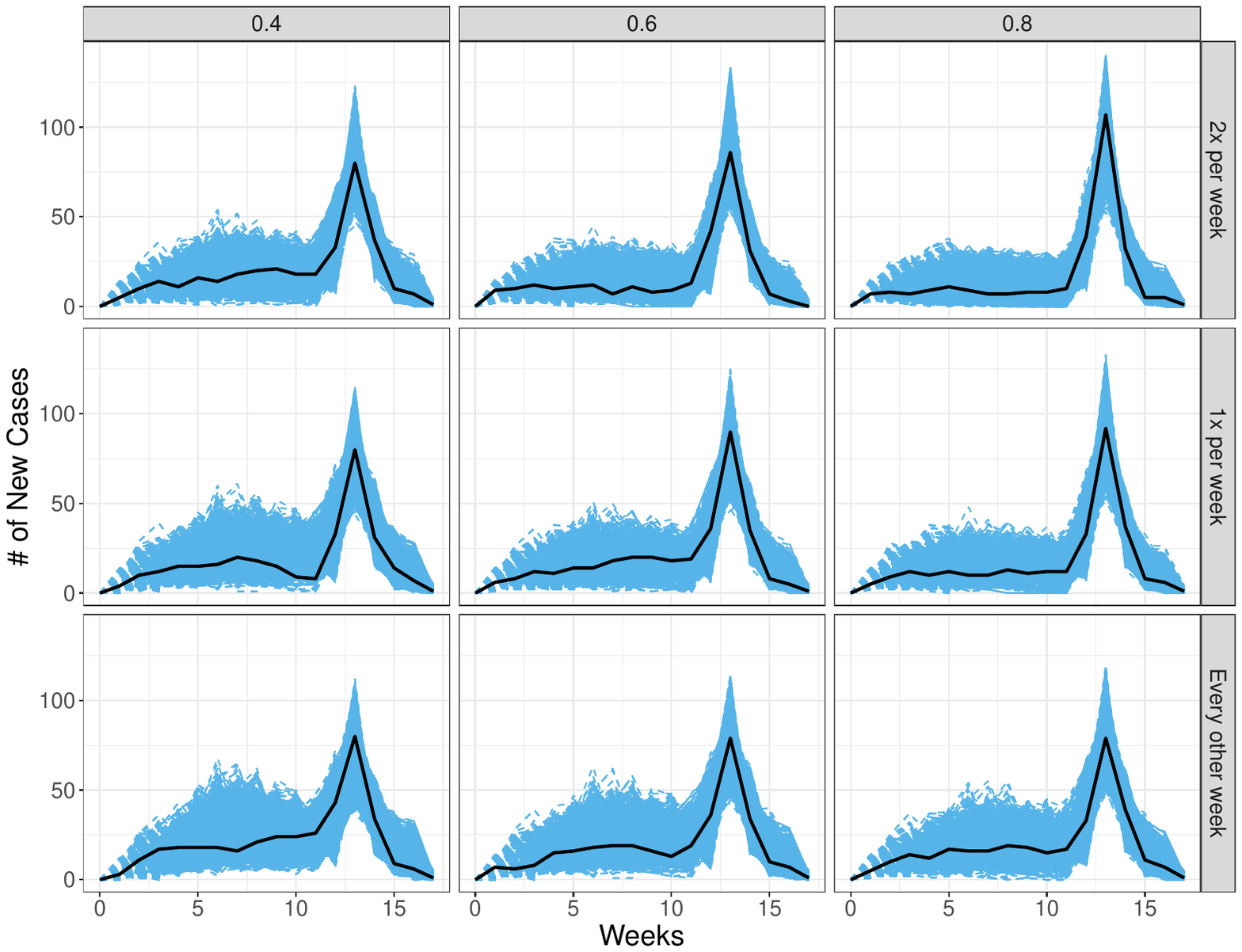}
    \caption{Simulated case trajectory curves for Fall 2020 under different testing scenarios. The black curve indicates the median in each scenario. The number of COVID-19 cases simulated from different testing frequencies and student testing percentages by day of the semester is shown. Adjustments in testing parameters show minimal influence on case trajectory curves. More testing leads to a slight increase in peak cases following fall break due to the detection of more cases. }
    \label{fig:new_cases_testing}
\end{figure}

Drawing samples from the posterior for our unknown parameters, we simulated the model under different potential surveillance testing strategies. Figure \ref{fig:freq_prop} illustrates the distribution of cases and students in quarantine at the end of the semester under various testing frequencies and percentages of students included in the surveillance testing program. An increase in both testing frequency and the number of students tested results in a reduction in reported cases. Increasing the number of students tested appears to be a slightly more efficient strategy. For example, testing 40\% of students twice a week yields a similar distribution of cases to testing 100\% of students every other week but utilizes 1500 less tests per week. Modifications to the testing strategy have only a marginal effect on the total number of students in quarantine by the end of the semester. This limited impact stems from our assumption that external infections will enter the campus following fall break, leaving insufficient time between the break and the semester's conclusion for a testing strategy to significantly mitigate this influx of cases affecting the number of students in quarantine.  In particular, more testing leads to a slightly higher peak in cases after the break because it results in the detection of more cases, as shown in Figure \ref{fig:new_cases_testing}. While more testing leads to a reduction of cases in the early part of the curves, adjustments in testing parameters have a minimal impact on the overall case trajectory curves. 

\section{Discussion}

We developed a model to simulate the spread of COVID-19, tailored to the dynamics specific to the Colorado School of Mines campus. Recognizing the inherent variability and multiple peaks in the observed case counts, we implemented a stochastic model to model transitions between different disease compartments. This decision to employ a stochastic model was driven by the inadequacy of a deterministic model in accounting for the observed variability in the data. As highlighted in \citet{kingAvoidableErrorsModelling2015}, stochastic models are favored over deterministic ones for their ability to better accommodate data variability and the uncertainties associated with model parameters. Additionally, they recommend models should be fit to disaggregated data whenever possible.  Thus, we tailored our model to the number of new cases reported per week rather than relying on the cumulative case curve. While the stochastic model provides a more accurate representation of COVID-19 dynamics on the Mines campus, it also poses challenges in parameter estimation. To address this, we developed an ABC method to estimate three key parameters in this model to campus case data and  draw the  posterior trajectories of the case counts.  To our knowledge this approach is new. With our calibrated model, we then explored the potential impact of different testing strategies that campus administrators might employ to limit the number of cases and students in quarantine throughout the semester

We used a comprehensive Monte Carlo simulation to  explore  different combinations of parameter values for: $\alpha, \beta$, and $I_{\text{out}}$. The ABC method on the simulated data yielded unbiased estimates of all three parameters. However, we observed that $\alpha$, representing the proportion of symptomatic cases, poses a particular challenge as it is not well identified from the data. In fact, the credible interval for $\alpha$ is nearly as broad as its entire possible range. Furthermore, the proportion of symptomatic infections ($\alpha$) is very sensitive to changes in the transmission rate ($\beta$).  The dependence between these two parameters is reasonable. Symptomatic individuals are likely to seek testing and therefore will transition to isolation faster than asymptomatic individuals. Thus, as the proportion symptomatic increases less spread will be occurring on campus. Therefore, the transmission rate will also need to increase to result in the same number of cases. Furthermore, we examined our simulation method under parameter values closely aligned with those derived from the true data. We found that the credible intervals closely resemble those obtained when using the actual data. This simulation study validates the robustness of our parameter estimation approach, affirming its effectiveness in accurately recovering the true values from our COVID transmission model.

We fit our model to the Fall 2020 campus case data, resulting in the determination of credible intervals for the unknown parameters. Similar to our simulation study, the credible interval for $\alpha$ exhibits substantial width. However, it's noteworthy that the range of possible values for all posterior parameters corresponds to estimated $R_0$ values falling within the range of 1.7 to 3.8. These estimates align with those from other studies of COVID-19 early in the pandemic \citep{katul2020global, d2020assessment, alimohamadi2020estimate}, .

We then employed our model under the estimated parameters to evaluate different testing policies that could be implemented on campus.  Increasing the proportion of students tested emerges as a slightly more efficient strategy,   reducing the number of cases during the semester while conserving total testing resources. However, changes in testing strategy have a limited effect on the peak number of students in quarantine. This limitation arises from our underlying assumption of an influx of cases into the campus after fall break,  thus increased testing led to more detected cases and consequently more students ending up in quarantine.  Furthermore, it's worth noting that an increase in testing leads to a slight rise in the peak number of cases following the break, a direct result of increased case detection. However, a silver lining exists in this scenario.  Increasing the number detection of cases in the population corresponds to a reduction in \textit{undetected} cases,  and contributes  to an ultimate reduction in the spread of the disease on a semester time scale.  

Our model presents several limitations. We assumed homogeneous mixing among the campus population, disregarding age-dependent transmission rates that could vary significantly between students and faculty. A more refined categorization into groups such as residents, non-residents, faculty, and staff could potentially offer a clearer picture of the disease dynamics among these cohorts. Additionally, the assumption of a closed campus population, while possibly appropriate during the early stages of the pandemic covered by this study, no longer reflects the current reality of increased student interactions with the broader community. Adapting the model to more accurately reflect the current situation could involve allowing the transmission rate ($\beta$) to vary dynamically, based on the infection rate within the surrounding community, as suggested in \citet{brownSimpleModelControl2021}. Although reinfection was not considered given the short time scale of our study, it would be essential to incorporate this possibility for modeling the disease spread in future semesters.  At the time of developing this model, a COVID-19 vaccine was not yet available. Future modifications could incorporate vaccination dynamics, including additional compartments and adjusted rates to account for vaccination, enhancing the model's relevance to ongoing public health efforts. 

The flexibility of our modeling and fitting approach ensures it can be adapted in response to changes in the pandemic's course and applied to other diseases with SEIR-type dynamics.  This framework not only advances understanding of COVID-19 dynamics on the Mines campus but also a blueprint for comparable settings, providing insights for informed strategies against infectious diseases. Successful implementation of the ABC method carefully tracking the context of the data analysis may encourage other applications of this approach for complex stochastic processes.

\section*{Acknowledgements}

Our sincere gratitude goes to the Office of Institutional Research at Mines for their generous support in funding this project. We extend our heartfelt thanks to Peter Han and the President's office at Mines for including us in this important work on campus and providing the insights that shaped our research questions. Special appreciation goes to Lisa Elson for her diligent efforts in gathering the numerous data sources necessary for this project. Additionally, we are grateful to the entire CRIT team at Mines for their hard work and invaluable assistance throughout the pandemic.

\section{Supplemental Materials}
\subsection{Next Generation Matrix calculation of $R_0$}

We have four ``infected" classes in the model. Let X be a vector of these infected classes and Y be a vector of the other (uninfected) classes:

$$X = \begin{bmatrix} E \\
I_A \\
I_{S} \\
I_{T}
\end{bmatrix}, \qquad \qquad   \qquad \qquad  \qquad \quad \qquad \qquad 
Y = \begin{bmatrix}
S \\
S_Q \\
Q_i \\
Q_Q \\
R_D \\
R_U
\end{bmatrix}.$$

\noindent Then,

$$ \frac{dX}{dt} = \mathcal{F}(X, Y) - \mathcal{V}(X, Y).$$

\noindent $\mathcal{F}(X, Y)$ is the vector of new infection rates (flows from Y to X). $\mathcal{V}(X, Y)$ is the vector of all other rates that don't create new infectious. In-flow in $\mathcal{V}$ is negative, out-flow is positive.

Define $ F = \begin{bmatrix}\frac{\partial \mathcal{F}}{\partial X}\end{bmatrix}$, and   $V = \begin{bmatrix}\frac{\partial \mathcal{V}}{\partial X}\end{bmatrix}$. The next generation matrix is defined as $FV^{-1}$. $R_0$ is equal to the maximum eigenvalue of the next generation matrix:
    


$$\mathcal{F} = \begin{bmatrix}
\beta S \frac{I_A + I_{S}+ I_T}{N}\\
0\\
0\\
0
\end{bmatrix}, \qquad \qquad   \qquad \qquad  \qquad \quad \qquad \qquad              
F = \begin{bmatrix}
0 & \beta  & \beta & \beta \\
0 & 0 & 0 & 0\\
0 & 0 & 0 & 0\\
0 & 0 & 0 & 0\\
\end{bmatrix},
$$

$$\mathcal{V} = \begin{bmatrix}
\mu E\\
\sigma \tau_f I_A + (1-\sigma)r_i I_A - (1- \alpha) \mu E\\
 \tau_S I_S - \alpha \mu E \\
\tau_r I_{t} - \sigma \tau_f I_A -  \tau_S I_S 
\end{bmatrix}, \;  \; \;     \;                          
V = \begin{bmatrix}
\mu & 0 & 0 & 0\\
-(1-\alpha)\mu & \sigma \tau_f + (1-\sigma)r_i & 0 & 0\\
-\alpha \mu & 0 &  \tau_S   & 0\\
0 & -\sigma \tau_f &  -\tau_S  & \tau_r\\
\end{bmatrix}.
$$

The maximum eigenvalue found in $FV^{-1}$ using \textsf{Mathematica} \cite{Mathematica} at the parameter values given in Table \ref{Tab:Parameters1}  is:

$$R_0 = \frac{ ( \alpha (\tau_S - \sigma \tau_f) \tau_R +  r_I \alpha(-1 + \sigma_A) (\tau_S + \tau_R) - \sigma_S (\sigma_A \tau_f + \tau_R)) \beta }{\tau_S \tau_R (r_I (-1 + \sigma_A) - \sigma_A \tau_F)},$$

$$ R_0 = (14.8 - 10.8\alpha)\beta.$$

The proportion of symptomatic individuals, $\alpha$, significantly influences the expected basic reproduction number, $R_0$. This is due to our assumption symptomatic individuals are more likely to get tested and therefore enter isolation, reducing their role in the transmission cycle. If all individuals were asymptomatic, $R_0$ would increase by a factor of 3.7 compared to a scenario where all individuals are symptomatic.






AI Statement: During the preparation of this work the authors used ChatGPT in order to edit and improve the readability of this work. After using this tool, the authors reviewed and edited the content as needed and take full responsibility for the content of the publication.

\bibliographystyle{abbrvnat}
\bibliography{bib}
\end{document}